\documentstyle[aps,prb,twocolumn,floats]{revtex}

\begin{document}

\newcommand{\stil}{\mbox{$\tilde{S}$}}  
\newcommand{\stiltwo}{\mbox{$\tilde{S}^2$}} 
\newcommand{\ktil}{\mbox{$\tilde{K}$}} 
\newcommand{\kper}{\mbox{$K_{\perp}$}}
\newcommand{\kpa}{\mbox{$K_{||}$}}
\newcommand{\ben}{\begin{equation}}
\newcommand{\een}{\end{equation}}
\newcommand{\as}{\mbox{$4 \tilde{S}^2$}}

\newcommand{\gsim}{
\,\raisebox{0.35ex}{$>$}
\hspace{-1.7ex}\raisebox{-0.65ex}{$\sim$}\,
}

\newcommand{\lsim}{
\,\raisebox{0.35ex}{$<$}
\hspace{-1.7ex}\raisebox{-0.65ex}{$\sim$}\,
}

\newcommand{\const}{ {\rm const} }
\newcommand{\arctanh}{ {\rm arctanh} }

\bibliographystyle{prsty}

\title{ \begin{flushleft}
{\small \em submitted to}\\
{\small 
xx
\hfill
VOLUME  XXX
NUMBER XXX
\hfill 
MONTH XXX
}\\
\end{flushleft}  
Phase transition between quantum and classical
regimes for the escape rate of a biaxial spin system
}      

\author{
Gwang-Hee Kim \cite{e-kim} 
}

\address{
Department of Physics, Sejong University,
Seoul 143-747, Republic of Korea\\
\smallskip
{\rm (Received xxx) }
\bigskip\\
\parbox{14.2cm}
{\rm
Employing the method of mapping the spin problem onto
a particle one, we have derived the particle Hamiltonian for a biaxial
spin system with a transverse or longitudinal magnetic field. 
Using the Hamiltonian and introducing the parameter 
$p ( \equiv (U_{\rm max}-E)/(U_{\rm max}-U_{\rm min}) )$ where
$U_{\rm max}$ ($U_{\rm min}$) corresponds to the top (bottom)
of the potential and $E$ is the energy of the particle, 
we have studied the first- or second-order transition
around the crossover temperature between 
thermal and quantum regimes for the escape rate,
depending on the anisotropy constant and the external magnetic field.
It is shown that the phase boundary separating the first-
and second-order transition and its crossover temperature
are greatly influenced by the transverse
anisotropy constant as well as the transverse or longitudinal magnetic field.
\smallskip
\begin{flushleft}
\end{flushleft}
} 
} 
\maketitle

In recent years, quantum-classical escape rate transition in the spin
system has emerged as good candidates to display first- or second-order
transition(FST).\cite{chu,gar,gmc,lmp}
Such a system is a single domain ferromagnetic particle with the
magnetization $\bf{M}$ whose direction is subject to the magnetocrystalline 
anisotropy. In this situation the direction $\hat{M}$ has at least two equivalent stable
orientations separated by an energy barrier $U$. Even though $\bf{M}$ is initially
directed along one of these equivalent orientations, $\hat{M}$ can be changed by
thermal activation whose rate is proportional to $\exp(-U/k_B T)$ at high
temperature, and by quantum tunneling at a temperature low enough to neglect
the thermal activation. In general, since the tunneling rate is dominated by
$\exp(-U/\hbar \omega)$ where $\omega$ characterizes the width
of the parabolic top of the barrier hindering the tunneling process, 
there is a crossover temperature $T_0$ from the thermally activated to
the quantum tunneling process. Whether the transiton about  the crossover   
temperature is first- or second-order one is determined by the external magnetic
field and the magnetic anisotropy constant. By controlling the field and choosing the
anisotropy constant in an appropriate way, we expect that there exists
the crossover temperature
$T^{(c)}_{0}$ at the  phase boundary between first- and second-order transition.

Theoretical studies of the transition have been around for some time. Affleck\cite{aff}
and Larkin and Ovchinnikov\cite{lar}
demonstrated that a second order transition from
thermal to quantum regimes can occur at the crossover temperature by using
the standard instanton technique. Later, Chudnovsky\cite{chu92}
discussed the criterion to
determine FST based on the behavior of the period of oscillations in the inverted
potential. Since then, based upon the mapping of a spin problem 
onto a particle one,\cite{zas}
Chudnovsky, Garanin and Mart\'\i nes\cite{chu,gar,gmc} suggested the spin
system with the uniaxial crystal symmetry which shows FST in the presence of a 
transverse and longitudinal field. A biaxial spin model without an applied field has
been considered by J.-Q. Liang {\it et al.}\cite{lmp}
who demonstrated that FST is determined
from the ratio of the transverse to the longitudinal anisotropy constant by using
the periodic instanton method. 
Even though they presented the analytical results without field, their approach
could not be simply extended to the situation in the presence of field.\cite{com1}
Very recently, in an effort to treat FST of the biaxial spin system with a longitudinal
field, Garanin and Chudnovsky\cite{gar98} used a perturbation approach and
obtained the phase boundary between the first- and the second-order transition which
is numerically corrected by the Liang {\it et al.}'s 
exact value in the absence of the field. Thus, 
relevant approach to treat a biaxial spin system with 
a transverse or longitudinal
field has been highly required for FST. In fact, 
complete analytical solution of the
problem seems to be considerablely important for the possibility of FST
in molecular magnetic systems as well as a single domain 
ferromagnetic particle with
many  spins. In this paper, employing the method of mapping
of a spin problem onto a particle one, we obtain the particle
Hamiltonian of the biaxial spin problem with a transverse or 
longitudinal field, which has not been obtainable from
the previous studies.\cite{lmp,chu92} Actually, such a mapping is
not a regular procedure and its form strongly 
depends on the form of the spin Hamiltonian.
Using the Hamiltonian and introducing the dimensionless energy
variable $p$, we will study the first- or second-order transition
around the crossover temperature between 
thermal and quantum regimes for the escape rate,
depending on the anisotropy constant and the external magnetic field,
and
present the analytic form of the phase boundary separating first- from
second-order transition and the crossover temperature at the 
phase boundary.

Consider the biaxial spin model in a transverse field $H_x$
described by the Hamiltonian
\ben
{\cal H} = -K_{||} S_z^2  + K_{\perp} S_y^2  - H_x S_x ,
\een
where $K_{||}$ and $K_{\perp}$ are the longitudinal and transverse
anisotropy constants, respectively.
This Hamiltonian can be mapped onto a particle problem\cite{zas}
which describes the exact
correspondence between the spin wave function 
\ben
\psi =\sum^{S}_{M=-S} a_M \left[ {(S+M)! (S-M)! \over (2 S)! } \right]^{1/2}
|S M \rangle,
\een
where $|S M \rangle$ are the eigenstates of $S_z$, and
the particle wave function 
$\Psi(x) = \exp(-\gamma(x)) \sum^{S}_{M=-S} a_M \exp(M x)$ where
\ben
{d \gamma(x) \over dx} ={{ \stil h_x  (1-k) \sinh(x)+(\stil-1) 
k \sinh (2 x) } \over {1+ k \cosh(2x)}},
\een
$\stil=S+1/2$, $k=k_t/(2+k_t)$, $k_t=\kper/\kpa$ and
$h_x =H_x /(2 \stil \kpa)$.
The particle Hamiltonian is 
\ben
{\cal H} =-{1 \over 2 m(x)}{d^2 \over d x^2 } +U(x),
\label{ham}
\een
where $1/m(x) =\kpa (2+k_t)  [1+k \cosh (2 x) ]$, $U(x)=\stiltwo \kpa u(x)$, and
\begin{eqnarray}
u(x)&=&  {1 \over (1-k) \left[1+k  \cosh (2 x) \right) }    \{ h^{2}_x \sinh^2(x) (1-k)^2  
\nonumber \\
&& -2 h_x (1-k)  \cosh(x) - k [2 h_x (1-k)  \cosh(x)
\nonumber \\
& & -1+\cosh(2x)-(5 \cosh (2x)-1)/(\as)]
\nonumber \\
&& + k^{2} [\cosh(2x)-1
\nonumber \\
&&+(\cosh^2(2x)-\cosh(2x)+4)/(\as)] \}.
\label{pot}
\end{eqnarray}
Here we notice that the potential and the mass 
without transverse anisotropy are
reduced to the results in Ref. \cite{zas}. Even though $\stil$ appears
in the potential, the terms associated with $\stil$ can be neglected in
the large $S$ limit including $S=10$. So, we will not include the terms
with $1/(\as)$ in the subsequent consideration.\cite{com2}

In the quasiclassical approximation the transition rate becomes
\ben
\Gamma \sim \int dE W(E) \exp[-(E-U_{\rm min})/T ],
\een
where $W(E)$ is the probability of tunneling at an energy $E$. 
Since this is defined via the imaginary-time action $W(E) \sim e^{-S(E)}$,
the transition rate is approximately given by
\ben
\Gamma \sim \exp(-F_{\rm min}/T),
\een
where $F_{\rm min}$ is the minimum of the effective ``free energy''
$F \equiv E+ TS(E)- U_{\rm min}$ with respect to $E$. Then, writing
$F(E)/T= \int^{1/2T}_{-1/2T} d\tau [ (m(x)/2) (dx/d\tau )^2 
+ U(x)-E_{\rm min} ]$,
the imaginary time action is given by\cite{chu,gar}
\ben
S(E) = 2 \int^{x_1(E)}_{-x_1(E)} dx \sqrt{2 m(x)} \sqrt{U(x)-E},
\label{act}
\een
where $\pm x_1(E)$ are the turning points for the particle with energy $-E$
in the inverted potential $-U(x)$. We note that the mass is coordinate dependent, which
is crucial in changing the  boundary between first- and second-order
transition, as will be seen. 
In order to determine FST in the crossover regime, we need to
consider the behavior of $S(E)$ near the top of the barrier. Since the potential
in Eq. (\ref{pot}) is even, we expand the integrand in Eq. (\ref{act}) near $x=0$
which corresponds to the top of the barrier. Introducing dimensionless energy
variable\cite{chu,gmc}
$p ( \equiv (U_{\rm max}-E)/(U_{\rm max}-U_{\rm min}) )$ where
$U_{\rm max}$ ($U_{\rm min}$) corresponds to the top (bottom)
of the potential, the action becomes near the top of the barrier
\ben
S(E)={ \pi \Delta U  \over \sqrt{U_2 (\kpa+\kper) } }  
[ p+\beta p^2+O(p^3) ],
\label{actg}
\een
where $U_2=-\tilde{S}^2 \kpa u^{(2)}(0)/2$, 
$U_4=\tilde{S}^2 \kpa u^{(4)}(0) /4!$, and
\ben
\beta={3 \over 8}\left( {U_4 \Delta U \over U^{2}_{2}} \right)
\left[ 1-{2 \over 3} \left( {k \over 1+k} \right) \times 
\left( { U_2 \over U_4} \right) \right].
\een
Here $\Delta U=U_{\rm max}-U_{\rm min}$ 
and the second term in the bracket
comes from the coordinate dependence of the mass. Using the
analogy with the Landau theory of phase transitions and
the general conditions for first- and second-order quantum-classical
transition of the escape rate discussed in 
Refs. \cite{lar} and \cite{chu92},
the factor in front of $p^2$ in Eq. (\ref{actg}) determines the  boundary
between the first- and second-order transition.

For the model without transverse field we have 
$U_2 =2 \tilde{S}^2 \kpa k/ (1+k)$ and
$U_4 =-2\tilde{S}^2 \kpa  k  (1-5k) / [3(1+k)^2] $
from Eq. (\ref{pot}).
If the mass does not depend on the coordinate, the sign of the
factor of $U_4$ determines whether the system becomes
the first- or the second-order transition due to the anharmonicity of
the potential. However, as has been already noticed from 
Eq. (\ref{ham}), the mass is a function of $x$. So, we cannot simply
obtain the  phase boundary from the anharmonicity of the potential
near the top of the barrier. Now, using 
$\Delta U= \stiltwo K_{||} $, the action (\ref{actg}) is given by
\ben
S(E)=(\pi \stil /\sqrt{k_t})  [p+ \beta p^2+ O(p^3)],
\een
where $\beta=(1-1 /k_t)/8$.
Thus the critical value of $k_t$ is $1$ implying that smaller values of
$k_t$ lead to the first-order transition which is consistent with the
result in Ref. \cite{lmp}.\cite{com3}

In case of the second-order transition the crossover occurs at
temperature $T^{(2)}_{0} =\omega_0 /(2 \pi)$  where $\omega_0$ is
oscillation frequency near the bottom of the inverted potential. To
estimate the frequency, we set up the Euclidean Euler-Lagrange
equation
$m(x) \ddot{x} +m^{\prime}(x) \dot{x}^2 /2 -U^{\prime}(x) =0$,
and insert $x=x_b +\delta x$  into the equation, 
where $x_b$ is $x$-coordinate of
the barrier. Expanding to second order in $\delta x$,
we have $\delta \ddot{x} +\omega^{2}_{0} \delta x =0$, where
$\omega_0 =\sqrt{|U^{\prime \prime}(0)|/m(0)}$. This yields the crossover
temperature given by
$T^{(2)}_0 = \stil  \sqrt{ \kpa \kper }/\pi$,
which leads to the crossover temperature at the  phase boundary,
$T^{(c)}_0 =\stil \kpa (2+k_t) /(3 \pi) $.

For the first-order transition the approximate form of  the
crossover temperature for small $k$
can be analytically calculated from the
relation $T^{(1)}_{0} \simeq \Delta U/S(E_{\rm min})$. At the bottom
of the barrier, $S(E_{\rm min})$ can be calculated directly from
the integral expression (\ref{act}) or following from\cite{lmp}
\ben
S(E)=4 \stil \sqrt{2 k -{\cal E} \over 1 -k}  [{\cal K}(q) -(1-\alpha^2)
\Pi (\alpha^2, q)],
\een
which is derived from the Eqs. (\ref{pot}) and (\ref{act}).
Here ${\cal E} = E/\stiltwo \ktil$, and ${\cal K}$ and $\Pi$ are
complete elliptic integrals of the first and third kind with
$q^2 =(z_1 -1)/(z_1 +1)$, $\alpha^2 = q^2 ( 1-k)/ (1+k)$, and
$z_1 =[k (k-1) +{\cal E}]/[k(k-1) - k {\cal E} ]$.
This yields
$T^{(1)}_{0} \simeq  (\stil \kpa /2)/ \ln \left[ (1+\sqrt{1-k^2}) / k \right]$,
which is approximate form for the first-order transition in the region of small $k$.

We now consider the model with a transverse magnetic field. 
Expanding the potential (\ref{pot}) in powers of $x$, we obtain
\begin{eqnarray}
U_2 &=&  [\stiltwo K_{||} /(1+k_t ) ] (h_x + k_t ) (1-h_x ) ,
\label{u2h} \\
U_4 &=&   [\tilde{S}^2 K_{||} /( 12 (1+k)^2  )] [ 4(1-5k) (1-k) h^{2}_{x}
\nonumber \\
& &+ (-1+38 k -57k^2) h_x - 8k(1-5k)  ],
\label{u3h} 
\end{eqnarray}
and $\Delta U=\stiltwo K_{||}  (1-h_x)^2$.
It is noted that the condition in which the barrier does not vanish is
$U_2 >0$, i.e., $h_x <1$.
Then, the action near the top of the barrier
becomes from Eq. (\ref{actg})
\ben
S(E)= \pi \stil { (1-h_x)^{3/2} \over (k_t+h_x)^{1/2} }[p+\beta p^2 +O(p^3)],
\een
where
\begin{eqnarray}
\beta&=&{(1-k_t) \over 8(h_x+ k_t )^2 } (h_x -h^{+}_{x})(h_x -h^{-}_{x}),\\
h^{\pm}_{x}(k_t) &=& {  1-14 k_t +k^{2}_{t} \pm(1+k_t) 
\sqrt{1+34 k_t+k^{2}_{t}}  \over 8(1-k_t )}.
\label{hpm}
\end{eqnarray}
As is shown in Fig. \ref{figh}, the behavior of $h^{\pm}_{x}$
shows that $\beta$ is negative for $k_t<1$ and $h_x <h^{+}_{x}$
which corresponds to the first-order transition. As is shown in 
Fig. \ref{figpha}, in the absence of the transverse anisotropy, $h_x=1/4$ is
the critical value for FST,\cite{chu} 
and in the presence of very small transverse 
anisotropy  we have 
$h^{+}_x \simeq (1+7 k_t /2)/4 $ in which
the  boundary becomes wider for $0 \leq k_t  < 0.2$.  
As $k_t$ continues to increase,
the region
where a first-order transition occurs, is smaller
for the biaxial model than for the uniaxial model.
This is intuitively understood that, since the transverse anisotropy
drives the decay of the metastable state, it plays the role of the transverse
field in the uniaxial case and so, for a given small transverse field the 
region for the first-order transition decreases as the transverse anisotropy
increases.
It is also noted that $h^{+}_x$ decreases linearly
for $k_t \lesssim 1$, i.e., just as $h^{+}_x \simeq (1-k_t)/3$. 

Continuing in the present case as without transverse case, we
have the crossover temperature for the second-order transition
\ben
T^{(2)}_{0}(k_t,h_x)= ( \stil \kpa/\pi ) 
 \sqrt{( k_t +h_x)(1-h_x)}.
\label{t02kh}
\een
Using Eq. (\ref{hpm}) for $h^{+}_{x}$, the crossover temperature at the
phase boundary between the first- and second-order transition is
written as
\ben
T^{(c)}_{0}= ( \sqrt{3} / (2 \pi) ) \stil \kpa 
( 1+k_t ) \sqrt{h^{+}_{x}(k_t) / (1-k_t )},
\een
which is illustrated in Fig. \ref{figtch}. We note that 
$T^{(c)}_{0} \simeq \stil K_{||} [\sqrt{3}/(4 \pi)] (1+3k_t)$ for
small $k_t$ and $T^{(c)}_{0} \simeq [\stil K_{||} /(2\pi) ] (1+k_t)$ 
for $k \lesssim 1$.

In case of the first-order transition the crossover 
temperature as a function of $k_t$ for small $h_x$ 
is approximately
estimated from
the ground-state tunneling exponent
given by Eq. (\ref{act}) or the direct integral
expression\cite{com4} and
$\Delta U$ considered above, which gives
\ben
T^{(1)}_{0} (k_t,h_x) \simeq  \stil \kpa (1-h_x)^2 / [2 
g(k_t, h_x) ] ,
\label{tc0h}
\een
where
\begin{eqnarray}
g(k_t,h_x)&=&\ln \left( { \sqrt{1+k_t}+\sqrt{ 1-h^{2}_{x} } \over
\sqrt{1+k_t}-\sqrt{ 1-h^{2}_{x} } } \right) 
- { 2 h_x \over \sqrt{k_t} }
\nonumber \\
& &  \times \arctan \left(
\sqrt{ k_t /( 1+k_t) } \sqrt{ 1/ h^{2}_{x} -1 }  \right).
\end{eqnarray}
Simple analysis for the crossover temperature
shows that there exists a maximum of the
crossover temperature $T^{\rm max}_{0}$ in the regime 
of the second-order
transition, which from Eq. (\ref{t02kh}) gives 
$T^{\rm max}_{0}= [\stil K_{||} /(2\pi) ] (1+k_t)$ at $h_x=(1-k_t)/2$.
It is noted that this 
is the asymtotic form of $T^{(c)}_{0}$ for $k_t \lesssim 1$ discussed
previously. As the transverse anisotropy increases, $h^{\rm max}_{x}$ for
the maximal crossover temperature decreases, (Fig. \ref{figpha}) while
$T^{\rm max}_{0}$ increases linearly. (Fig. \ref{figtch})
As is summarized in Table \ref{tabsum} and shown in Fig. \ref{figtch},
as $k_t$ increases, the difference between $T^{\rm max}_{0}$ and
$T^{c}_{0}$ decreases and becomes zero at $k_t=1$ which is
the critical value in the fieldless case.

In the presence of a longitudinal field of the spin model
${\cal H} = -K_{||} S_z^2  + K_{\perp} S_y^2  - H_z S_z$,
we can proceed the consideration, similarly
and so, 
we will  briefly discuss  the essential points.
In order to obtain the relation between $h_z[=H_z/(2 \stil \kpa)]$ and
$k_t$ at the  phase boundary, we need to have the coefficient of
$p^2$ in the action, i.e., $\beta$ including 
$U_3$ and $m^{\prime}(x_b)$ as well as $U_2$ and $U_4$
where $x_b$ is the position of the barrier, because the potential
given by
\begin{eqnarray}
u(x)&=& {1 \over  (1-k) \left[ 1+k  \cosh (2 x) \right]  } \{ h^{2}_z (1-k)^2
 \nonumber \\
&-& k [2 h_z (1-k)  \sinh(2x) -1+\cosh(2x)
\nonumber \\
&-&(5 \cosh (2x)-1)/(\as)]
\nonumber \\
&+& k^{2} [\cosh(2x)-1
\nonumber \\
&+&(\cosh^2(2x)-\cosh(2x)+4)/(\as)] \},
\label{pothz}
\end{eqnarray}
is not an even function. Following the procedure discussed previously,
we have the  boundary between first- and second-order transition 
\ben
k_t =(1-h^{2}_{z})/ ( 1+ 2 h^{2}_{z} ).
\label{kt}
\een
The ratio of two anisotropy constans, $k_t$ decreases parabolically
for $h_z \ll 1$,
as $k_t \simeq 1- 3 h^{2}_{z}$, and linearly for
$h_z \lesssim 1$,
as $k_t \simeq (2/3)(1-  h_z)$. This can be understood from
the fact that, since
the height of barrier decreases 
as $h_z$ increases, the first-order transition is
expected for the larger width of the barrier which comes from
the smaller value of the trasverse anisotropy.
The crossover temperature 
$T^{(c)}_0$ at the phase
boundary can be obtained by using Eq. (\ref{kt}) and
$T^{(2)}_{0}(k_t,h_z) =(1 /2 \pi) \sqrt{|U^{\prime \prime}(x_b)|/m(x_b)}$
in the second order transition, which leads to
\ben
T^{(c)}_{0} =(\stil \kpa/\pi)(1-h^{2}_{z})/\sqrt{1+2 h^{2}_{z}}.
\een
Simple analysis shows that our results are consistent with the ones
deduced from the correction of the perturbative calculation performed
in Ref. \cite{gar98} up to the numerical factors. 
Even though the
perturbative approach is less justified at large value of the 
transverse anisotropy constant, its  boundary and
crossover temperature are strikingly the same trend as
our analytical results obtained from the quasiclassical method. 

In this paper we have considered the quantum-classical escape rate
transition of a biaxial spin system in the presence of a transverse
or longitudinal field by using the particle Hamiltonian mapped from the spin
system. The coordinate dependence of the particle
mass was crucial in changing the  boundary between the first- and
second-order transition and its boundary was greatly influenced by
the transverse anisotropy constant and external field, whose effect is expected
to observe in future experiment including ${\rm Fe_{8}}$ molecular
magnet.\cite{san}.

I am indebted to E. M. Chudnovsky for many useful discussions.
This work was supported
by the Basic Science Research Institute Program,
Ministry of Education, Project No. BSRI-98-2415.\\

\begin{table}
\caption{
Summary of (a) the maximum crossover temperature and
(b) the crossover temperature at the  phase boundary
at two end points in Fig. \protect\ref{figtch}.
}
\label{tabsum}
\end{table}

\begin{center}
\begin{tabular}{|c|c|c|}  \hline \hline
 $k_t$ & $h_x$ & $T_0 /(\stil \kpa/2\pi)$ \\
\hline 
0 & 1/2 & $1 \ \ ({\rm a} )$ \\
1 & 0 & $2 \ \ ({\rm a} )$ \\
0 & $h^{+}_{x}=1/4$ & $\sqrt{3}/2 \ \ ({\rm b} )$ \\
1 & $h^{+}_{x}=0$ & $2 \ \ ({\rm b} )$ \\
\hline \hline
\end{tabular}
\end{center}

\begin{figure}
\caption{
$h^{-}_{x}$ as a function of $k_t$. Inset: $h^{+}_{x}$ as 
a function of $k_t$.
}
 \label{figh}
\end{figure}

\begin{figure}
\caption{
Boundary between the first- and the second-order transitions,
where (a) $h^{+}_{x}(k_t)$,
(b) $h^{\rm max}_{x}(k_t)$ which gives the maximal crossover
temperature in the quantum-classical transition, and (c)
the boundary whether the barrier vanishes or not.
}
\label{figpha}
\end{figure}

\begin{figure}
\caption{
Dependence of (a) the crossover temperature $T^{(c)}_{0}$ at
the  phase boundary and (b)
the maximum crossover temperature $T^{\rm max}_{0}$
on the scaled anisotropy constant $k_t$ 
where $\bar{T}^{(c)}_{0} =T^{(c)}_{0} /(\stil \kpa /2\pi) $.
}
\label{figtch}
\end{figure}


\begin{thebibliography}{10}
\bibitem{com2} We check this point numerically and confirm that, up the
small numerical values which can be neglected, there
is no difference of $u(x)$ between with and without $1/(\as)$-terms in the
integration range from $-x_1$ to $x_1$ in Eq. (\ref{act}).
\bibitem{com3} Since the Hamiltonian 
${\cal H}=K_1 S^{2}_{z}+K_2 S^{2}_{y}$ 
is used in Ref. \cite{lmp}, our anisotropy
constants are related by $K_{||}=K_2$ and $K_{\perp}=K_1-K_2$.
This yields $k_t=-1+1/\lambda$, in which $k_t=1$
corresponds to $\lambda=1/2$.
\bibitem{com4} We have also performed the calculation for
the action in the whole range of $E$. In this case the formula is
very complicated and we do not present them here. We checked
that the formula in the bottom of the potential agrees with
the result (\ref{tc0h}) obtained from the action at the ground-state
tunneling.
\end{thebibliography}
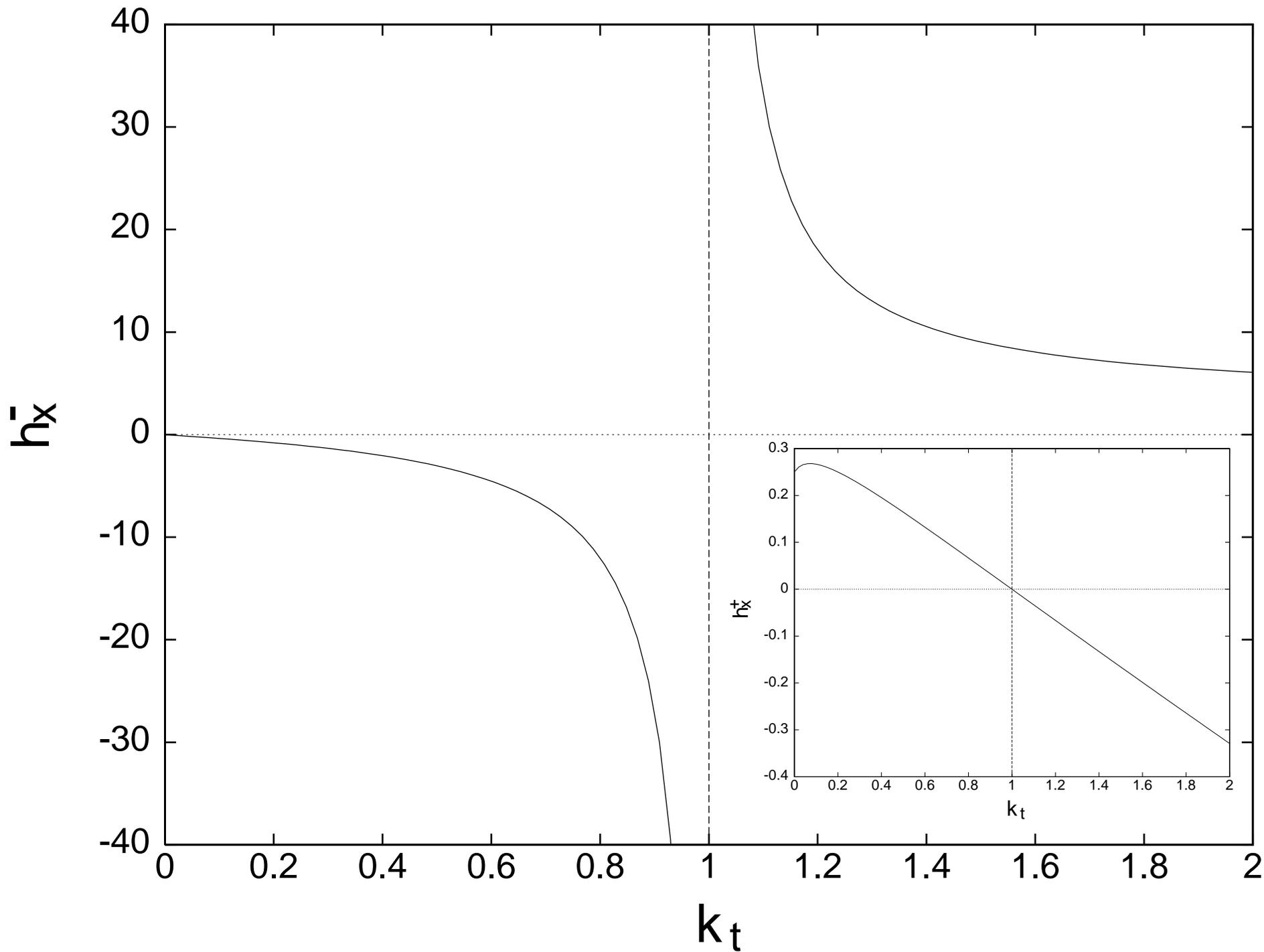
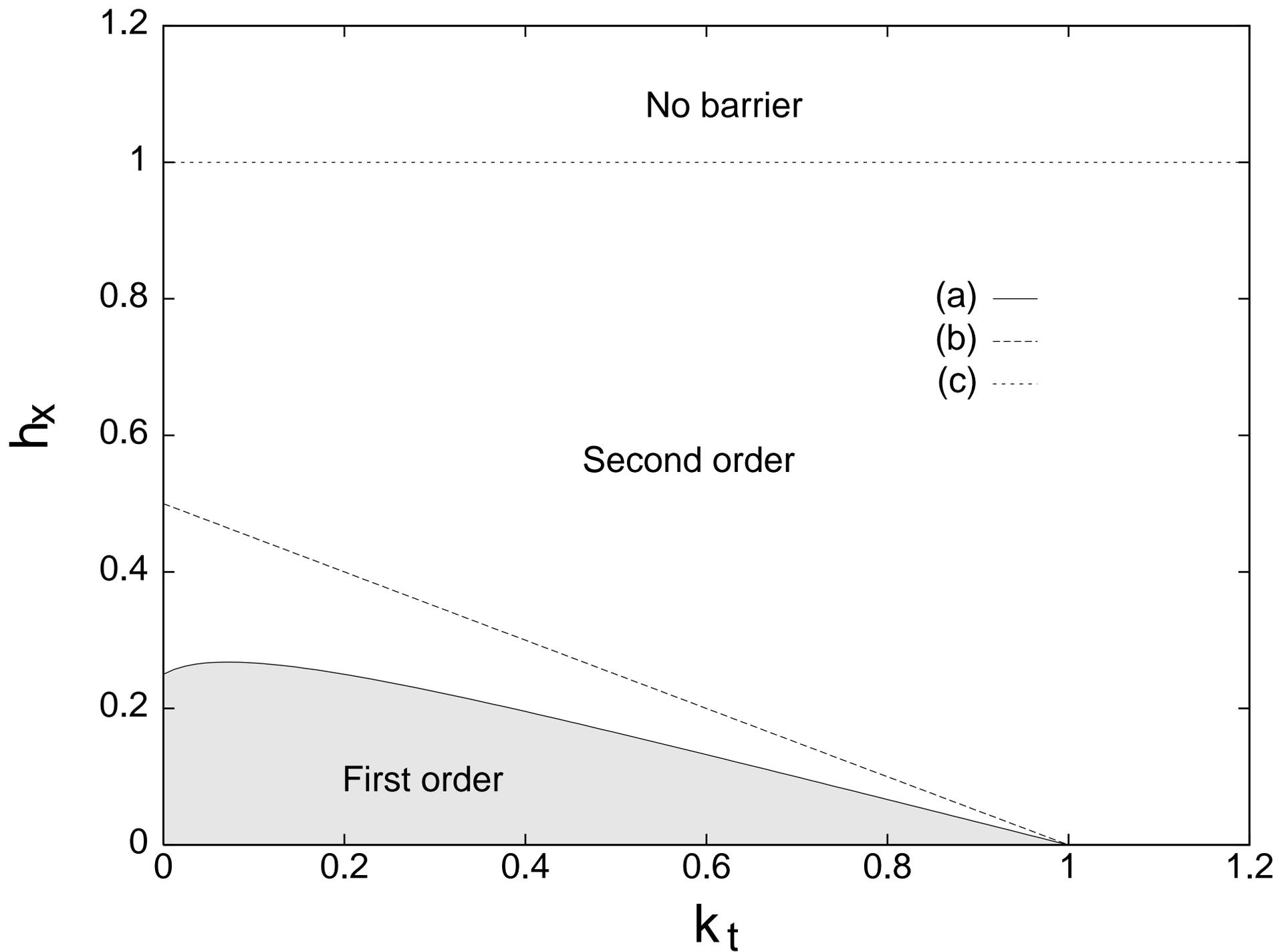
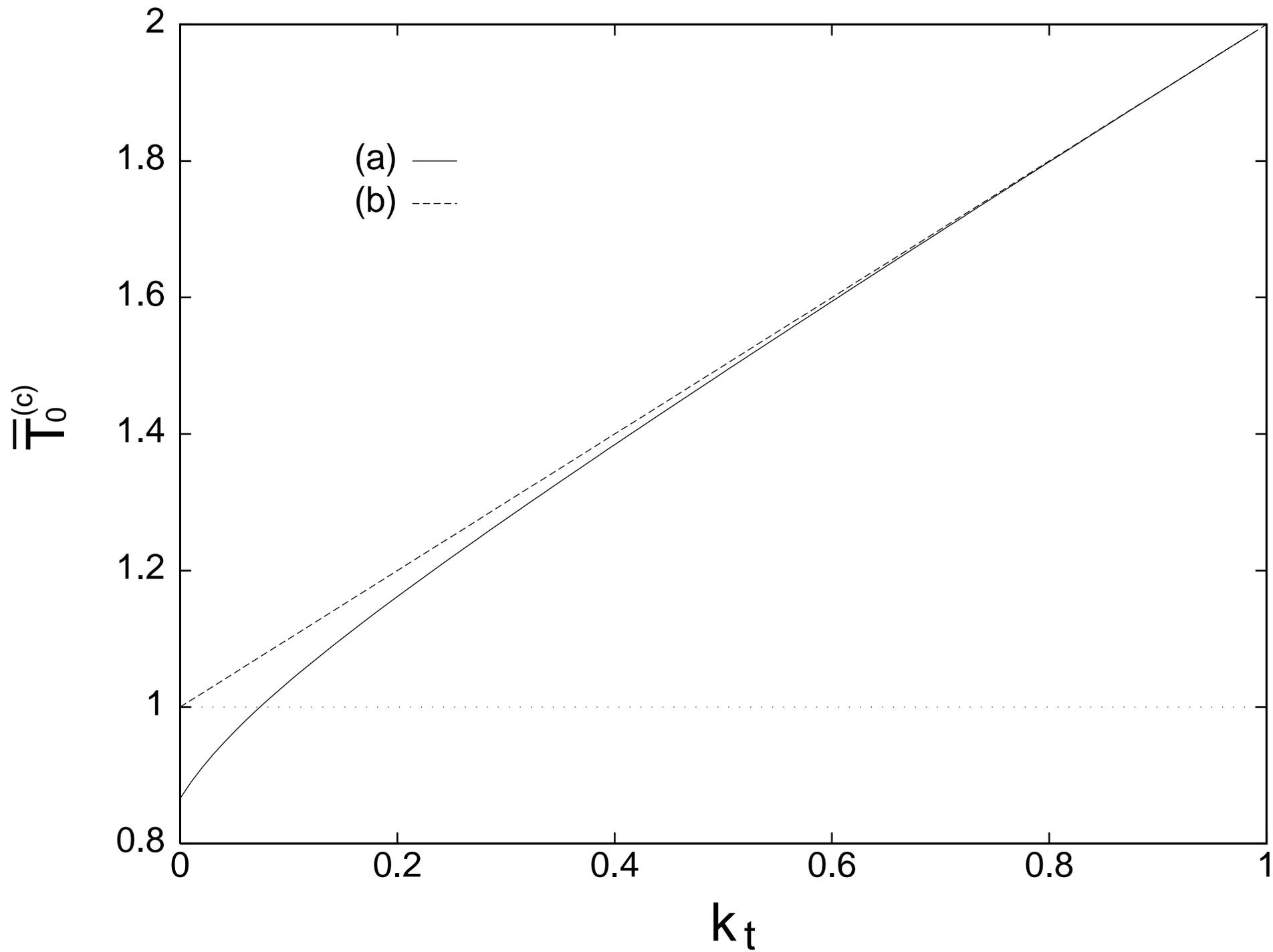
\end{document}